\documentclass[twocolumn,prl,showpacs,floatfix]{revtex4}

\usepackage{graphicx}

\begin{document}

\title
{
Is relaxation to equilibrium hindered by transient dissipative 
structures in closed systems?
}
    
\author
{ 
Akinori Awazu and Kunihiko Kaneko
}

\affiliation
{
Department of Pure and Applied Sciences, University of Tokyo, Komaba 3-8-1, 
Meguro-ku, Tokyo 153-8902, Japan.
}

\date{\today}

\begin{abstract}
Dissipative structures are generally observed when a system relaxes from 
a far from equilibrium state. To address the reverse question given by 
the title, we investigate the relaxation process in a closed chemical
reaction-diffusion system which can potentially form Turing-like patterns 
during the transient. We find that when certain conditions are fulfilled 
the relaxation process is indeed drastically hindered, once the pattern is 
formed. This slowing-down is shown to be due to stepwise relaxation, where 
each plateau in the relaxation process corresponds to residence at a 
certain spatial pattern. Universality of the phenomena as well as their 
biological relevance is briefly discussed. 
\end{abstract}

\pacs{}

\maketitle

In open systems far from equilibrium, organized structures are well-known
phenomena\cite{pri0}. Such ``dissipative structures'' include temporal 
rhythms and spatial patterns in chemical reaction-diffusion systems, 
hydrodynamical systems, optical systems and so 
forth\cite{pri0,haken,kuramo,pearson}. 
Among others, biological complex structures can also be maintained by 
non-equilibrium conditions\cite{tokei,burumen}.

In the study of dissipative structures, systems are generally prepared in 
far from equilibrium states by imposing certain constraints. For example, 
the concentrations of some chemicals are fixed at (compared to the 
equilibrium state) higher or lower levels by supplying or removing them 
from the outside. On the other hand, in biological systems, non-equilibrium 
conditions are maintained autonomously, at least when considering long
time spans. As a first step for understanding the autonomous sustainment 
of biological non-equilibrium conditions, it is of interest to investigate 
the possibility that the longevity of the conditions for the dissipative 
structures is extended by the formation of the structures themselves.

In closed systems, of course, equilibrium states without any structures 
are reached eventually. However, oscillatory behaviors or spatial pattern 
formations can be observed as transient phenomena during the course of 
relaxation to equilibrium\cite{e1,e3,e4,ri1,ri11,ri2,ri3}. Here, we 
address the following question: Can the formation of (transient) 
dissipative structures make far-from-equilibrium conditions last 
significantly longer by slowing down the relaxation process to equilibrium?  
To answer this question, we study the relaxation behaviors of a closed 
coupled chemical reactor that can potentially form transient Turing-like 
patterns during the relaxation process.

Here, in contrast to most studies in reaction-diffusion systems, we need 
to take the changes in the concentrations of all chemicals into account, 
instead of keeping the concentrations of some chemicals constant. Thus we 
consider the following reaction-diffusion system consisting of the reactions 
I)$A+v+2u \rightleftharpoons^{k_{AB}}_{k_{BA}} B+3u$, 
II)$u \rightleftharpoons v$, and III)$A+u \rightleftharpoons A+v$, and 
also diffusion. Considering  the limiting case $k_{AB}= 1 >> k_{BA}$, 
the evolution of the chemical concentrations is given by
\begin{equation}
\dot{u}_{i} = A_{i} v_{i} u_{i}^{2} -(1+A_{i})(u_{i} -v_{i}),
\end{equation}
\begin{equation}
\dot{v}_{i} = -A_{i} v_{i} u_{i}^{2} -(1+A_{i})(v_{i} -u_{i}) +D_{v}(v_{i+1}+v_{i-1}-2v_{i}),
\end{equation}
\begin{equation}
\dot{A}_{i} = -A_{i} v_{i} u_{i}^{2} +D_{A}(A_{i+1}+A_{i-1}-2A_{i}).
\end{equation}
Here, $u_{i}$, $v_{i}$, and $A_{i}$ denote the concentrations of the chemical 
components and $i$ denotes the index of each site in a one-dimensional space, 
where periodic boundary conditions are adopted for $i$.  Each chemical
diffuses to neighboring sites with a diffusion coefficient $D_X$ 
($X=u,v, or A$). Although we adopt a spatially discrete system for simplicity,
the conclusions drawn do not change even when the continuum-limit 
(partial differential equation) is taken. The diffusion coefficient 
$D_u$ is assumed to be slow, and we mostly study the case with $D_{u}=0$ 
as in eq(1) since this will not affect our findings qualitatively as long 
as $D_{u} << D_{v}$. The chemicals $u_{i}$, $v_{i}$ and $A_{i}$ can be 
regarded as activators, inhibitors and resources respectively.

This model is a variant of the Gray-Scott\cite{ri11,pearson} or
Brusselator\cite{pri0,ri2} models such that we additionally include
 changes in the resources $A_{i}$.  Note that the value 
$\frac{1}{2N}\sum_{i}(u_{i}+v_{i}) = S$ is conserved due to the system 
being closed ($N$ is the number of sites.). 

While relaxation to a unique equilibrium state satisfying $A=0$ and 
$u_{i}=v_{i}=S$ is assured for $t \to \infty$ in this model, if we fix 
$A_{i} = Ao >> 1$ in order to maintain the non-equilibrium condition,
this system shows the following bifurcation of the attractor, depending on 
$S$. I) If $S \le 0.75$, a unique uniform state with $u_{i}$ and $v_{i}$ 
constant over $i$ and time exists that is stable against small perturbations. 
II) If $S > 0.75$, the uniform state is unstable, and the attractor is 
replaced by a non-uniform pattern of $u_{i}$ and $v_{i}$, which is constant 
in time. This Turing instability of the uniform states is straightforwardly 
obtained by linear stability analysis. 

In order to study the relaxation process from a non-equilibrium state to the
homogeneous equilibrium state, we now investigate the effects of changing
$A_{i}$ as in eq.(3) for the case $S > 0.75$. Here, depending on the initial 
configurations of $u_{i}$, spatial patterns can be formed during relaxation 
to the homogeneous equilibrium state.
We study typical relaxation behaviors by varying initial configuration of 
$u_{i}=u_{i}^{0}$ with $S=4$ and the initial condition $A_{i}=A_{ini}=100$ 
and $v_{i}=S=4$. We control initial spatial inhomogeneity of $u_{i}^{0}$ by
taking an initial condition $S + \delta \times rnd_{i}$, with $rnd_{i}$ as a
uniform random number over [-1,1] ($0 \le \delta \le S$).

Two sets of typical temporal evolutions of $u_{i}$ and $A_{i}$ are
displayed in Figs. 1(a) and 1(b) where $\delta=0.1$ in (a) and $\delta=4.0$ 
in (b) with $D_{v}=250$ and $D_{A}=0$. The pattern is plotted until it has 
nearly reached the equilibrium state. The corresponding time evolution of 
$<A>$ ($= \frac{1}{N} \sum_i^N A_{i}$) is plotted in Fig. 1(c).

When $\delta$ is small, $u_{i}$ remains almost flat with only minor
fluctuations, and no structure is formed as in Fig. 1(a). In this case $<A>$
decreases smoothly with time as indicated by the solid curve in Fig. 1(c).
On the other hand, when  $\delta$ is large, the initial inhomogeneity in 
$u_{i}$ is amplified leading to the formation of a nonuniform spatial 
pattern that is sustained over some time span until it is re-organized into 
a different pattern, as shown in Figs. 1(b) and (d).  In this case, the 
relaxation of $<A>$ exhibits some plateaus as shown in Fig. 1(c), and 
requires much more time as compared to the case when $\delta$ is small. 
In general, several plateaus are observed during the relaxation process, 
each of which corresponds to a specific spatial pattern, as shown in 
Fig.1(d).

Hence, we have found an explicit example in which the formation of 
a dissipative structure slows down the relaxation process. This behavior is 
rather general in our model, as long as $S$ and $D_v$ are large enough to 
allow for the formation of spatial patterns.

In order to obtain insight into the relationship between pattern and 
relaxation, we have measured the spatial inhomogeneity of $u_{i}$ defined by 
$F(t)= \frac{1}{N} \sum_{i}^{N}|u_{i+1}-u_{i}|^{2}$. In Figure 2 we plot the 
decay rate of $<A>$ defined by $<A>'=\frac{d log<A>}{dt}$, as a function 
of $F(t)$. As can be seen, the system alternates between structure 
formation where $A_{i}$ is consumed and residence at the formed 
non-uniform structure where consumption of $A_{i}$ is suppressed.  
Indeed, the decrease of $<A>'$ is highly correlated with the increase of $F$. 
Thus, the slowing down of the relaxation process by the spatial structure is 
confirmed.

Next, we study the conditions for this slowing down of the relaxation process.
We investigate the dependence of the relaxation time on the initial 
inhomogeneity. Figure 3(a) shows the sample average of the relaxation time 
$T$ as a function of the initial heterogeneity $\delta$, computed up to the 
time when $<A>$ has decreased to $0.1A_{ini}$. Here, the parameters are set 
to $D_v=250$, $S=4$, $A_{ini}=100$ and $N=200$, while the diffusion constant 
$D_{A}$ is chosen to be $0$, $0.25$, and $25$. As can be seen, there is a 
critical inhomogeneity $\delta_c$ ($\approx 0.5$), beyond which the 
relaxation time increases, when $D_{A}$ is small. Indeed, $\delta_c$ is 
nothing but a threshold for the inhomogeneity above which the the 
re-organization of the spatial structure is possible. For large $D_{A}$, 
however, the re-organization of the structure is even then not possible, and 
the relaxation time is insensitive to the initial fluctuations. The threshold 
$\delta_c$ exists for $D_{A} \stackrel{<}{\approx} 20$, while the value 
of $\delta_c$ itself is insensitive to the value of $D_{A}$ within this 
range.

The dependence of the relaxation time on $D_{A}$, on the other hand, is
plotted in Figure 3(b) where $\delta=0.1$ or $=4$.  When $\delta$
is smaller than $\delta_c$, the relaxation time remains short.  
In contrast, the relaxation time shows a peak
around $D_{A} \sim 0.3$, when $\delta$ is larger than $\delta_c$,
while for smaller and larger $D_{A}$, it approaches constant
values. Here the relaxation time for smaller $D_{A}$ remains large,
while it is quite small for larger $D_{A}$, as already shown in
Fig. 3(a).

Now, we study the mechanism for the observed drastic enhancement of the
relaxation time, by exploring the mechanism behind the dependence of the
relaxation process on the spatial fluctuations in the case of
small $D_{A}$ as in Fig. 1. If $u_{i}$ is distributed almost
uniformly, $A_{i}$ is consumed almost uniformly through the entire
system as in Fig. 1(a). On the other hand, if $u_{i}$ has large
spatial variations, $A_{i}$ is consumed locally at the sites with large
$u_{i}$, resulting in an amplification of the spatial fluctuations, as
shown in Fig. 1(d). Then, at sites with large $u_{i}$, $A_{i}$ is
consumed rapidly so that the reaction 
$A+v+2u \rightleftharpoons^{k_{AB}}_{k_{BA}} B+3u$ hardly occurs at these
sites ($k_{AB} =1 >> k_{BA}$.). The decrease of $u_{i}$ there 
mainly progresses by the reaction $u_{i} \rightleftharpoons v_{i}$
which does not consume $A_{i}$.

On the other hand, with this spatial structure, at several sites 
that do not belong to the peaks in the pattern, $u_{i}$ and $v_{i}$ are 
suppressed. Then, the reaction between $u$ and $v$ with the consuming $A$ 
(whose rate is $A_{i}v_{i}u_{i}^{2}$) is highly suppressed, even if 
$A_{i}$ therein is large. $A_{i}$ at such sites with small $u_{i}$ is 
consumed only little by little.  Hence, a plateau appears in the relaxation 
of $<A>$.
After the decay of $u_{i}$ at the site with large $u_{i}$ is completed, 
$u_{i}$ at some other site that keeps large $A_{i}$ starts to be amplified by
consuming $A_{i}$. This process corresponds to the re-organization of
spatial structure of $u_{i}$ as described in Fig. 1(b) and Fig. 1(d).
In this way, several plateaus appear successively during the
relaxation process.

Based on the above argument, the drastic increase in the relaxation time by
the increase of the initial heterogeneity $\delta$ is explained
by comparing the speed of the reaction consuming $A_{i}$ with the reaction
$u_{i} \rightleftharpoons v_{i}$. Note that for a stationary state, 
$v_{i} u_{i}^{2} \sim u_{i}-v_{i}$ is satisfied (by disregarding diffusion) 
if $A_{i} >> 1$.  When the locally conserved quantity $u_{i}+v_{i} =2S$ is 
large, this stationary solution satisfies $v_{i} \sim 1/u_{i}$. Then, the 
consumption speed of $A_{i}$ can be approximated by 
$\sim A_{i}v_{i}u_{i}^{2} \sim A_{i}u_{i}$, according to the estimate for 
the steady state. On the other hand, the order of the speed of the reactions
$u_{i} \rightleftharpoons v_{i}$ is given by $\sim u_{i}$, which is much 
smaller when $<A> \gg 1$. Accordingly, the relaxation process of $<A>$ with 
the re-organizations of $u_{i}$ involve fast and slow
processes, with reaction speeds $\sim A_{i}u_{i}$ and $\sim u_{i}$.
Thus, the relaxation of $<A>$ has several bottlenecks corresponding to the
slow processes with speeds in the order of $\sim u_{i}$, leading to plateaus.
In contrast, the relaxation from a state with little inhomogeneity
involves only a fast process with a reaction speed $\sim A_{i}u_{i}$,
since the inhomogeneity in $u_{i}$ is not amplified to form a structure.
Hence, the relaxation time is small when  $\delta< \delta_c$.

Next, we explain the $D_{A}$-dependence of the relaxation time in 
Fig. 3. As $u_{i}$ increases at a certain site $i$, the resource $A$
diffuses into this site from adjacent sites as $A_{i}$ is consumed at site 
$i$. Thus the increase of $u_{i}$ is further accelerated by the use of 
$A_{j}$ at the sites adjacent to $i$. Consequently, the peak height of 
$u_{i}$ can be much higher than is the case when $D_{A}=0$. With such 
increases of the spatial amplitudes, the time intervals between 
re-organizations of $u_{i}$ become larger, resulting in an increase of 
the relaxation time.

On the other hand, if $D_{A} >> 1$, the resource $A$ is consumed
faster due to the diffusion of $A$. In this case, the speed of the flow
of $A_{i}$ is higher than that due to the reaction 
$u_{i} \rightleftharpoons v_{i}$ for sites with large $u_{i}$. Therefore, 
the resource $A_{i}$ is consumed continuously by sites with large
$u_{i}$, and a re-organization of the spatial structure $u_{i}$ no longer
occurs.  In this case, the consumption speed of $A_{i}$ goes up to the
level for the relaxation from a homogeneous pattern.  Hence, the
relaxation time to equilibrium for the case $D_{A} >> 1$ is much
smaller than for the case with $D_{A} < 1$, even when $\delta>\delta_c $. The
peak of the relaxation time in Fig. 3(b) is thus explained.

Note that the mechanism for the slowing down of relaxation processes 
proposed here is rather general. Take any reaction-diffusion system in 
which dissipative structures are formed by constraining the concentrations 
of some resource chemicals in such a way that their values are
larger than their equilibrium values.  Then consider
the corresponding closed system, where the dynamics of the
resource chemical(s) is incorporated.  Now the proposed mechanism for
slowing down the relaxation process
is possible if the following two conditions are fulfilled.

First, the reaction and diffusion processes of the chemicals
that give rise to the non-equilibrium conditions should not be too
fast compared to those of the other chemicals. The diffusion constants
as well as the reaction rates for the consumption of the resource chemicals 
should be smaller than those for the others. This leads to
differences in the time scales of the concentration changes and thus, when the
pattern formation progresses fast enough, the consumption of resource 
chemicals to support the non-equilibrium conditions slows. 
Then, the changes of resource chemical concentrations work as a kind of slow
variables (or parameters) of the system.

Second, the consumption of resources should slow down due to feed-back
from the spatial structure. In the present example, the consumption of 
resources is completed soon at sites with higher $u_i$ and for other sites 
with very small concentrations of $u$, the consumption reaction progresses 
only slowly. Hence the overall depletion of resources is much slower than 
in the case with a homogeneous concentration of $u$.
 
In general, it is not particularly difficult to satisfy these two 
conditions, and indeed we have confirmed the present mechanism by studying 
some variants of the present reaction-diffusion system.

If only the second condition is satisfied but not the first condition,
some increase in the relaxation time is still observed, but it is neither 
drastic nor does the relaxation have several plateaus.  As an example we 
consider the reaction system i)$a_{i}+v_{i}+2u_{i} \rightleftharpoons^{1}_{k}
a_{i}'+3u_{i}$, ii)$b_{i}+u_{i} \rightleftharpoons b_{i}+v_{i}$,
iii)$e_{i}+u_{i} \rightleftharpoons e_{i}+c_{i}$, iv)$f_{i}+v_{i}
\rightleftharpoons f_{i}+d_{i}$, with diffusion.  If we set $e_{i}=0$
and $c_{i}=c (=const)$, this model is equivalent to a 
Brusselator\cite{pri0,ri2} model with the reactions reversed, while
it corresponds to the Gray-Scott model\cite{ri11,pearson}, again with the
reactions reversed, if $b_{i}=0$ and $d_{i}=d (=const)$. In this case too, 
Turing patterns are formed if the concentrations of resource and waste 
chemicals are fixed as $a_{i} = a_{o}$ with $k << 1$ in order to keep the 
non-equilibrium condition. By including the dynamics of $a_{i}$ and by 
choosing a large $a_{i}$ initially, a structure is formed if the initial 
inhomogeneity $\delta$ is not too small, in the same way as for our model 
above. Indeed, some amplification of the relaxation time is observed here.  
However, once the initially formed structures are destroyed, new large 
structures are not produced. Hence, the enhancement of the relaxation time 
in this case is not as significant as in the previous case.

In this Letter, the relaxation process to equilibrium is investigated 
through a closed coupled chemical reactor system. Under certain conditions, 
we have found that the relaxation is drastically hindered once a pattern 
is formed. In addition, we have observed repeated formations of patterns. 
With this itinerancy over patterns, the relaxation is further slowed down 
as compared to the case without the structure formation. 

In experimental studies of dissipative structures, the system under 
consideration is usually set to be open in order to sustain the 
non-equilibrium condition. Still, even in closed systems, dissipative 
structures are often observed as transients which may last for rather long 
time spans (recall for example the Belousov-Zhabotinsky reaction in a 
petri dish). By choosing a suitable reaction system, it will be possible to 
demonstrate the present enhancement of the relaxation time due to the 
transient dissipative structure.
In relationship with the possibility of an experimental observation, it 
should be noted that the slow relaxation process we studied here does not 
progress gradually, but repeats residence at several quasi-stationary states 
corresponding to certain structures. The quasi-stationary states associated
with the successive spatial structures act as bottlenecks for the
relaxation process.

In complex reaction systems, with more chemical components, the relaxation 
process could further be slowed down.  For example, assume that $A$ and $B$ 
in our model are synthesized by lower-level resources $A'$ and $B'$, and that
these reactions also satisfy the mechanism demonstrated in the present 
paper. By a hierarchy of such reactions, the relaxation time is then expected 
to be further increased. This may provide some insight into why a cell 
system can maintain a non-equilibrium state over a huge time span.

The authors are grateful to F. H. Willeboordse for useful 
discussions. This work is supported by Grant-in-Aids for Scientific Research 
from the Ministry of Education, Science and Culture of Japan (11CE2006).

\begin{figure}
\begin{center}
\includegraphics[width=8cm]{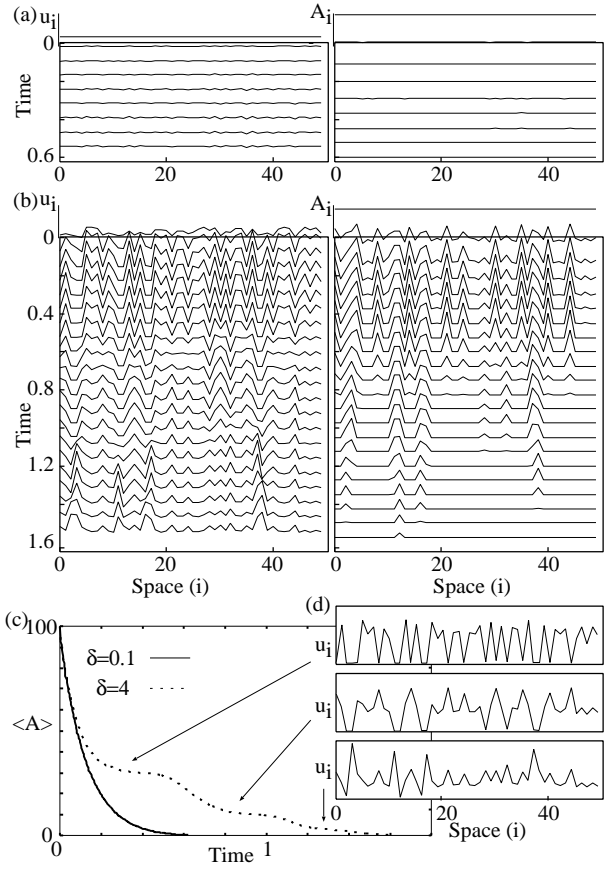}
\end{center}
\caption{Typical temporal evolutions of $u_{i}$ (left), $A_{i}$ (right) for 
$D_v=250$, $D_A=0$, and $S=4$. (a)$\delta=0.1$ and (b)$\delta=4.0$ plotted 
until $<A>$ becomes smaller than $0.001A_{ini}$.
(c) The time evolutions of $<A>$ corresponding to 
(a) and (b). (d) Three typical snapshots of spatial patterns in (b) 
which are plotted at the time step shown by the arrows.}
\end{figure}

\begin{figure}
\begin{center}
\includegraphics[width=6cm]{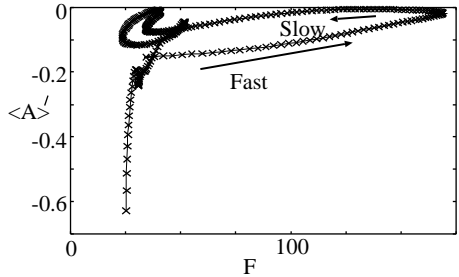}
\end{center}
\caption{Time course of $(<A>'(t),F(t))$  obtained from the same simulation 
of Fig. 1(b). See text for the definition of $<A>'$ and $F$.}
\end{figure}

\begin{figure}
\begin{center}
\includegraphics[width=8cm]{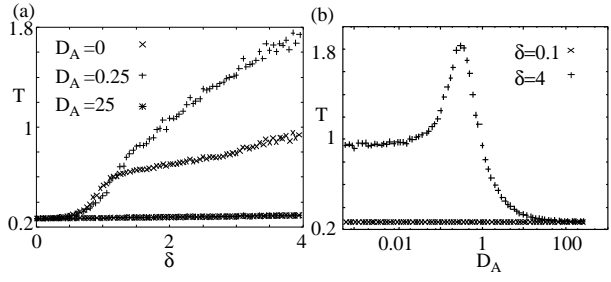}
\end{center}
\caption{(a) The average relaxation time $T$, plotted as a function of 
$\delta$, for $D_{A}=0$, $=0.25$, and $=25$, and (b) $T$ as a function of 
$D_{A}$ for $\delta=0.1$ and $=4$. $D_v=250$.}
\end{figure}

\end{document}